\documentclass[10pt, conference]{IEEEtran}
 \IEEEoverridecommandlockouts

\usepackage{cite}

\ifCLASSINFOpdf
\else
\fi
\usepackage{color}
\usepackage{graphicx}
\usepackage[colorlinks,
            linkcolor=red,
            anchorcolor=blue,
            citecolor=green
            ]{hyperref}
\usepackage{breakurl}   
\usepackage{subfig}
\usepackage{fixltx2e}
\usepackage{listings}
\usepackage{fixltx2e}
\definecolor{lightgray}{rgb}{.9,.9,.9}
\definecolor{darkgray}{rgb}{.4,.4,.4}
\definecolor{purple}{rgb}{0.65, 0.12, 0.82}

\lstdefinelanguage{JavaScript}{
  keywords={typeof, new, true, false, catch, function, return, null, catch, switch, var, if, in, while, do, else, case, break},
  keywordstyle=\color{blue}\bfseries,
  ndkeywords={class, export, boolean, throw, implements, import, this},
  ndkeywordstyle=\color{darkgray}\bfseries,
  identifierstyle=\color{black},
  sensitive=false,
  comment=[l]{//},
  morecomment=[s]{/*}{*/},
  commentstyle=\color{purple}\ttfamily,
  stringstyle=\color{red}\ttfamily,
  morestring=[b]',
  morestring=[b]"
}

\newcommand{\code}[1]{\texttt{\small#1}}

\begin{document}
%
\title{Opportunities in Software Engineering Research \\ for Web API Consumption}

\author{\IEEEauthorblockN{Erik Wittern*, Annie Ying*, Yunhui Zheng, Jim A. Laredo, Julian Dolby, Christopher C. Young\textsuperscript{+}, Aleksander A. Slominski}
\IEEEauthorblockA{IBM T. J. Watson Research Center, Yorktown Heights, NY, USA\\
Email: \{witternj, aying zhengyu, laredoj, dolby, ccyoung, aslom\}@us.ibm.com}
\thanks{* First authors}
\thanks{\textsuperscript{+} Currently at Harled Inc in Kitchener, ON, Canada. Email: chris@harled.ca}
}


\maketitle


\begin{abstract}

Nowadays, invoking third party code increasingly involves calling web services via their web APIs, as opposed to the more traditional scenario of downloading a library and invoking the library's API.  
However, there are also new challenges for developers calling these web APIs.
In this paper, we highlight a broad set of these challenges and argue for resulting opportunities for software engineering research to support developers in consuming web APIs.
We outline two specific research threads in this context:
(1) web API specification curation, which enables us to know the signatures of web APIs, and
(2) static analysis that is capable of extracting URLs, HTTP methods etc. of web API calls.
Furthermore, we present new work on how we combine (1) and (2) to provide IDE support for application developers consuming web APIs.
As web APIs are used broadly, research in supporting the consumption of web APIs offers exciting opportunities.
\end{abstract}


%
\IEEEpeerreviewmaketitle


\section{Introduction}
\label{sec:intro}
Programmers write applications using a growing variety of publicly accessible web services, or by interacting with dedicated, private backends.
Applications typically consume both, these services and backends, using \emph{web APIs} -- application programming interfaces invoked over network that rely on web technologies like HTTP as a transport protocol or XML and JSON as data formats.
In practice, these APIs are often "REST-like", in that they adhere to some of the constraints imposed by the \emph{Representational State Transfer} (REST) architectural style~\cite{Fielding:2000}.
The extent of the proliferation of web APIs is indicative in the thousands of web APIs listed in catalogs such as IBM's API Harmony~\cite{APIHarmony}, Mashape's PublicAPIs~\cite{Mashape}, or ProgrammableWeb~\cite{ProgrammableWeb}.

For developers, consuming web APIs poses many challenges, compared to that of traditional library APIs, including:

\begin{itemize}
  \item \textbf{Challenge 1:} Clients have the guarantee that once they download a traditional library, they have control over the API and the code being called. However, web APIs clients have no control over the API and the service behind the API, as a provider may change either or both.
  \item \textbf{Challenge 2:} Clients of a local library can depend on a compiler to check whether a call conforms to the library interface. Clients of web APIs, in contrast, do not know whether the signature of a call---i.e., the URL, the request payload, query parameters---is valid until run-time.\footnote{Though, there are frameworks for remote procedure calls, e.g., Thrift~\cite{slee2007thrift}.}
  \item \textbf{Challenge 3:} Even when clients invoke a web service via a \emph{software development kit} (SDK), the SDK is essentially a wrapper that ultimately sends requests via the web API. A web API call in an SDK can get out of synchronization with the actual web API, a even more likely concern with third-party SDKs.
  \item \textbf{Challenge 4:} Invoking web services opens a range of issues concerning remote calls, including asynchrony, service availability, service latency and more generally, quality of service (QoS) issues. Handling these issues may require additional client code and/or restructuring of the architecture.
\end{itemize}
These and similar challenges have rarely been addressed in software engineering research, especially comparing to the vast amount on research on more traditional library APIs~\cite{robillard2013automated}.
In our opinion, the lack of research to support the consumption of web APIs is surprising, given the broad usage of web APIs.

In this paper, we argue for research in software engineering to support consumption of web APIs.\footnote{There are also many opportunities to support the provisioning of web APIs, which is not the focus of this paper.}
In section~\ref{sec:vision}, we outline our vision of such research and state concrete research opportunities.
Next, we present two examples of such research efforts.
First, in section~\ref{sec:spec}, we discuss work on web API specification curation, which enables us to know the signatures of web APIs~\cite{Suter:2015}.
Second, in section~\ref{sec:analysis}, we discuss work on static analysis that is capable of extracting web API calls from source code, including URLs, HTTP methods, and request data~\cite{Wittern:2017}.
Furthermore, in section~\ref{sec:atom}, we demonstrate the practical significance of such research by presenting a novel IDE integration tying together these two efforts, for warning a programmer of possible errors in the web APIs calls in their JavaScript code, directly addressing the aforementioned Challenge 2. Section~\ref{sec:conclusion} concludes.


\section{Our vision in Web API Research}
\label{sec:vision}
We envision that a first line of research relates to static analysis and IDE support for developers writing code containing web API calls. Some examples include the following:

\begin{itemize}
  \item Detect errors in source code invoking web APIs, e.g., by statically checking that requests (comprising possibly the URL, HTTP method, query parameters, a request payload) match APIs' requirements (our work~\cite{Wittern:2017}, describe in Section~\ref{sec:analysis}). Provide developer support in IDEs, e.g., for error checking (Section~\ref{sec:analysis}) and auto-completion.
  \item Provide refactoring support, for example to help developers to migrate to new web API versions, similar to efforts for traditional APIs, e.g.,~\cite{dagenais2011recommending,wu2010aura}.
  \item Mine web API usage for more advanced developer support, e.g., recommending effective web API usage patterns, provisioning code snippets or even recommending effective composition of APIs, similar to such efforts for traditional library APIs, for which Robillard et al. provides an extensive survey~\cite{robillard2013automated}.
\end{itemize}

Another line of research involves mechanisms to document a web API's signatures; interfaces of web APIs are not automatically available to clients, unlike in traditional library APIs where the API is part of a library's signatures.  The API's signatures are needed, e.g., for the aforementioned error checking and version migration. 
Machine-readable API specifications have been proposed, like the \emph{OpenAPI specification}~\cite{OAI}.  However, these specifications are not necessarily available.  A promising research area is as follows:

\begin{itemize}
  \item Support the curation of web API specifications, e.g., the OpenAPI specification~\cite{OAI}. 
  Research can address how to automatically create, maintain, or test such specifications (Section~\ref{sec:spec}).
\end{itemize}

Because of the nature of web API calls, clients need additional software engineering support, posing opportunities of research, including:
\begin{itemize}
  \item Research coding practices and patterns for an application to deal with varying QoS, especially if used from different geographic locations~\cite{Bermbach:2016}.  
  \item Assess the impact of web API usage on non-functional aspects.  E.g., network usage is found to be the major consumer of battery life in mobile applications~\cite{Li:2014}, and security and privacy concerns raised by the usage of external services are of high priority to users~\cite{Chin:2012}.
\end{itemize}

To reap such opportunities, researchers in this new field of supporting web API consumption have many technical opportunities in addressing web APIs.
They include requests being made asynchronously over the network, mostly using untyped and stringified data (e.g., our static analysis in Section~\ref{sec:analysis}), the lack of web API specifications (e.g., our work in automatically generating specifications in Section~\ref{sec:spec}), or that web APIs are commonly used from dynamic languages, e.g., JavaScript, a primary language supported by browsers.  


\section{Curation of Web API Specifications}
\label{sec:spec}
A first line of research addresses the automatic creation and maintenance of web API specifications, which are a required input for many mechanisms to support web API consumption. As we will also do in Section~\ref{sec:analysis}, we start by motivating this work, then summarizing our research in this area, and providing advice for how such research can be evaluated.

\subsection{Problem Motivation}
\label{sec:spec_motivation}
Web APIs do not per default expose descriptions of their interfaces to clients (cf. Challenge 2 in Section~\ref{sec:intro}). Rather, developers have to familiarize with individual endpoints using human-readable documentation. The lack of machine-readable interface descriptions limits the creation of capabilities like static request checking (as we will see in Section~\ref{sec:analysis}), auto-completion, refactoring support, automatic testing, or API composition support.

To address this shortcoming, various web API specification or description formats exist, like the aforementioned OpenAPI specification~\cite{OAI} (previously known as \emph{Swagger}), the \emph{web application description language} (WADL)~\cite{WADL}, or the \emph{RESTful API Modeling Language} (RAML)~\cite{RAML}. These formats describe aspects concerning the whole API like authentication or security, as well as information about individual endpoints, including the data required to send in requests and the responses to expect. Despite the existence of API specification formats, they are rarely available to developers in practice, especially for public APIs. Third-party efforts that create and maintain specifications of popular APIs like APIs Guru~\cite{APIsGuru}, which rely on API-specific mining facilities and contributions from the open source community, exemplify this sparsity.

The problem this line of research is concerned with is to automate the creation and maintenance of such specifications.

\subsection{Challenges and Possible Solutions}
\label{sec:spec_challenges}
A straight-forward solution to address the lack of API specifications is to create them as part of web API provisioning efforts. Indeed, \emph{API management} solutions that aim to help providing APIs like IBM's API Connect~\cite{APIConnect} rely on having a specification upfront or entering a comparable set of information through graphical user interfaces (from which later a specification can be exported). 
Similarly, web API specifications can be created using source code annotations, e.g.~\cite{SwaggerCore,SwaggerJSDoc}. 
However, both approaches require manual effort (especially for legacy APIs) for initially creating \emph{and} for maintaining specifications. Also, they depend upon API providers spending efforts, as consumers have no access to the API source code or operations.

To automate specification creation and maintenance, research has looked into inferring specifications from dynamic server traces~\cite{Suter:2015} or from observed HTTP requests~\cite{Sohan:2015b}. Beyond these efforts, possible directions to look into include inferring specifications from human-readable documentation, or from API usage mined from open source repositories (cf. Section~\ref{sub:analysis-evaluation}). If working, these approaches would require less or even no manual effort, and could be used to keep specifications in sync with API implementations.
However, sparse input data is a main challenge for such approaches. Usage data, for example, only helps to learn about parts of web APIs that are actually used, leaving the risk of incomplete specifications. If requests are encrypted, request and response payloads are not available. Also, many web APIs feature endpoints that can best be described with path templates. In them, dynamic URL segments that depend on user input or runtime parameters are explicitly denoted, for example \texttt{username} in \texttt{.../user/\{username\}/profile}. However, input data like dynamic traces only contains concrete request instances like \texttt{.../user/erik/profile}. The resulting challenge, thus, is to infer more generic API specifications from (few) examples. An example on how to address this challenge is our previous work on inferring specifications from dynamic traces, where we attempted solving this problem using machine-learning techniques~\cite{Suter:2015}.

\subsection{Evaluation and Relevance to Web API Researchers}
\label{sec:spec_evaluation}
To evaluate the curation of web API specifications, publicly available specifications can act as ground-truth, e.g.~\cite{APIsGuru}. As further input data, web API usages can be obtained from open source repositories like GitHub. Human-readable API documentation pages are made available by most large API providers. In our previous work, we obtained real-world usage of IBM's Watson APIs for one month and compared inferred specifications against the human-written specifications of said APIs~\cite{Suter:2015}. Specifications are central to many other research opportunities, amplifying the importance of this line of research.


\section{Static Analysis on Web API Calls}
\label{sec:analysis}
A second line of research addresses the lack of support for statically checking web API calls in JavaScript code~\cite{Wittern:2017}.

\subsection{Problem Motivation}
\label{sub:analysis-motivation}
\begin{figure}
  \centering
  \includegraphics[width=\columnwidth]{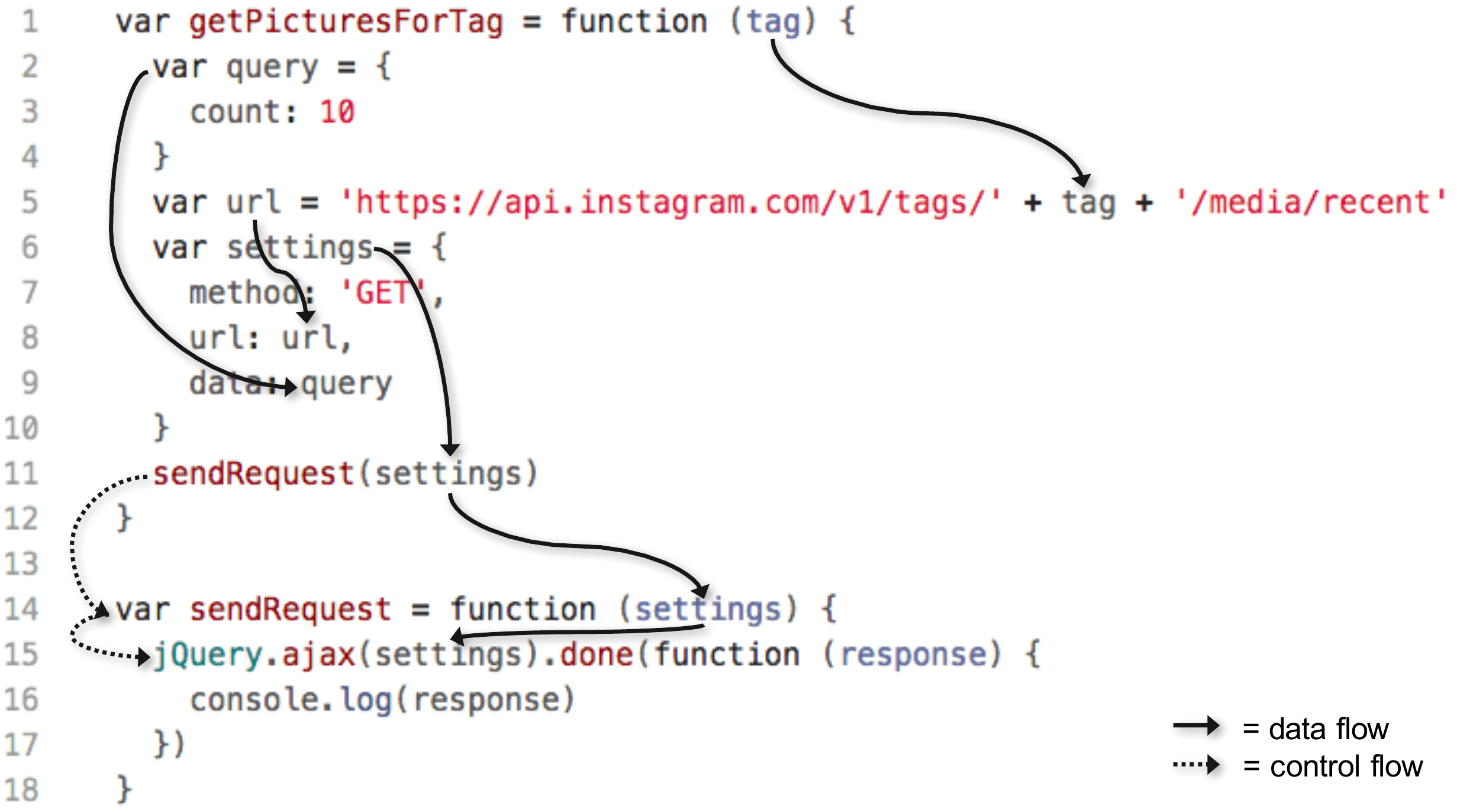}
  \caption{Code excerpt of a request to the Instagram API}
  \label{fig:example_inter_procedual}
  \vspace{-0.1in}
\end{figure}

To invoke a web API, applications can send HTTP requests to a dedicated URL using one of its supported HTTP methods; required data is sent as query or path parameters, or within the HTTP request body. The URL, HTTP method, and data to send are strings possibly constructed by string operations within the applications.
Figure~\ref{fig:example_inter_procedual} shows an exemplary excerpt of a JavaScript application performing these actions.
When a request targets a URL that does not exist or sends data that does not comply with the requirements of the web API, a run-time error occurs. This prevalent calling mechanism for web APIs does not allow type-safety checking. In other words, checks for traditional compile-time errors are not available for programmers writing code calling web APIs. 

\subsection{Our Approach and Relevance to Web API Researchers}
\label{sub:analysis-details}
To address this problem, we are working on a static checker that takes as input OpenAPI specifications, and then statically checks whether the web API requests in JavaScript code conform to these specifications. 
This solution embodies two technical challenges relevant to other researchers in the web API reseach field:
\begin{itemize}
  \item Our approach has to be able to extract requests which are strings (i.e., the URL string, HTTP method, and the corresponding data from a request), using an inter-procedural static string analysis (i.e., program analysis capable of extracting strings~\cite{Feldthaus_icse2013}).
  In comparison, static analysis on traditional library API calls in a statically-typed language can be extracted simply via an abstract syntax tree.
  \item Because web APIs are commonly used in code written in dynamically typed languages such as JavaScript, we targeted our analysis on JavaScript code. Typically, inter-procedural static whole-program analysis on JavaScript code is more challenging (i.e., less scalable, due difficulties from dynamical typing) than the same analysis on Java code~\cite{DBLP:conf/pldi/SchaferSDT13,DBLP:conf/oopsla/AndreasenM14,DBLP:conf/kbse/KoLDR15,Feldthaus_icse2013}.
  For the initial implementation, we chose to handle requests written using the jQuery framework due to its popularity -- reportedly, ~70\% of websites use the jQuery framework~\cite{jQueryUsage}.
\end{itemize}

\subsection{Evaluation and Relevance to Web API Researchers}
\label{sub:analysis-evaluation}
To evaluate our approach, we applied our approach by checking whether web API requests from over $6000$ JavaScript files on GitHub\footnote{\url{https://github.com/}} were consistent or inconsistent with publicly available web API specifications provided by the APIs Guru project~\cite{APIsGuru}. 
Other researchers in web API research may also want to consider using the public web API specifications such as the data from APIs Guru project for the evaluation of approaches like our static analysis.


\section{IDE Support for Checking Web API Requests}
\label{sec:atom}
This section demonstrates the practical significance of research we presented in Sections~\ref{sec:spec} and~\ref{sec:analysis} through a new IDE integration taking advantage of both research efforts.

From a high level, providing static checking comprises two steps:
First, we use inter-procedural static analysis~\cite{Wittern:2017} discussed in section~\ref{sec:analysis} to extract string-based information about the request, including the request URL string, HTTP method, and the corresponding request data. Recall the code in Figure~\ref{fig:example_inter_procedual} as an example. Focusing on the \code{url} variable, we can see that it is composed from two constant strings and the \code{tag} variable in the function \code{getPictureForTag}. The value of \code{url} is then first passed into a field of the \code{settings} object, and ultimately flows into jQuery's Ajax call \code{\$.ajax} in the \code{sendRequest} function. The value of \code{tag} is a parameter and could be different in multiple runs. Hence, when we aim to extract the URL used in this request, we denote \code{tag} as a symbolic value \code{\{tag\}} using curly braces, indicating that the value is not known statically. Overall, the URL extracted for the shown request is \code{https://api.instagram.com/v1/tags/ \{tag\}/media/recent}.

Second, a checking component determines whether extracted requests conform to any known web API specification, possibly curated as described in section~\ref{sec:spec}. To do so, the checking component retrieves one or more specifications from the database that define a host matching the one in the extracted request URL. Multiple specifications may be found as multiple versions of an API can share the same host. The component then in sequence checks whether the protocol, base path, route, and HTTP method match. If any of these matches fail for \emph{all} of the retrieved specifications, an error is reported. If at least one endpoint (route and HTTP method~\cite{Suter:2015}) was matched across the specifications, the component checks if required query parameters are part of the extracted URL and if payload data requirements are met by extracted payload data. Again, if any of these conditions fail for all matched endpoints, an error is reported. Otherwise, the request is deemed valid. This approach is conservative in that it only reports errors if no match can be found at all.

\begin{figure}
  \centering
  \includegraphics[width=\columnwidth]{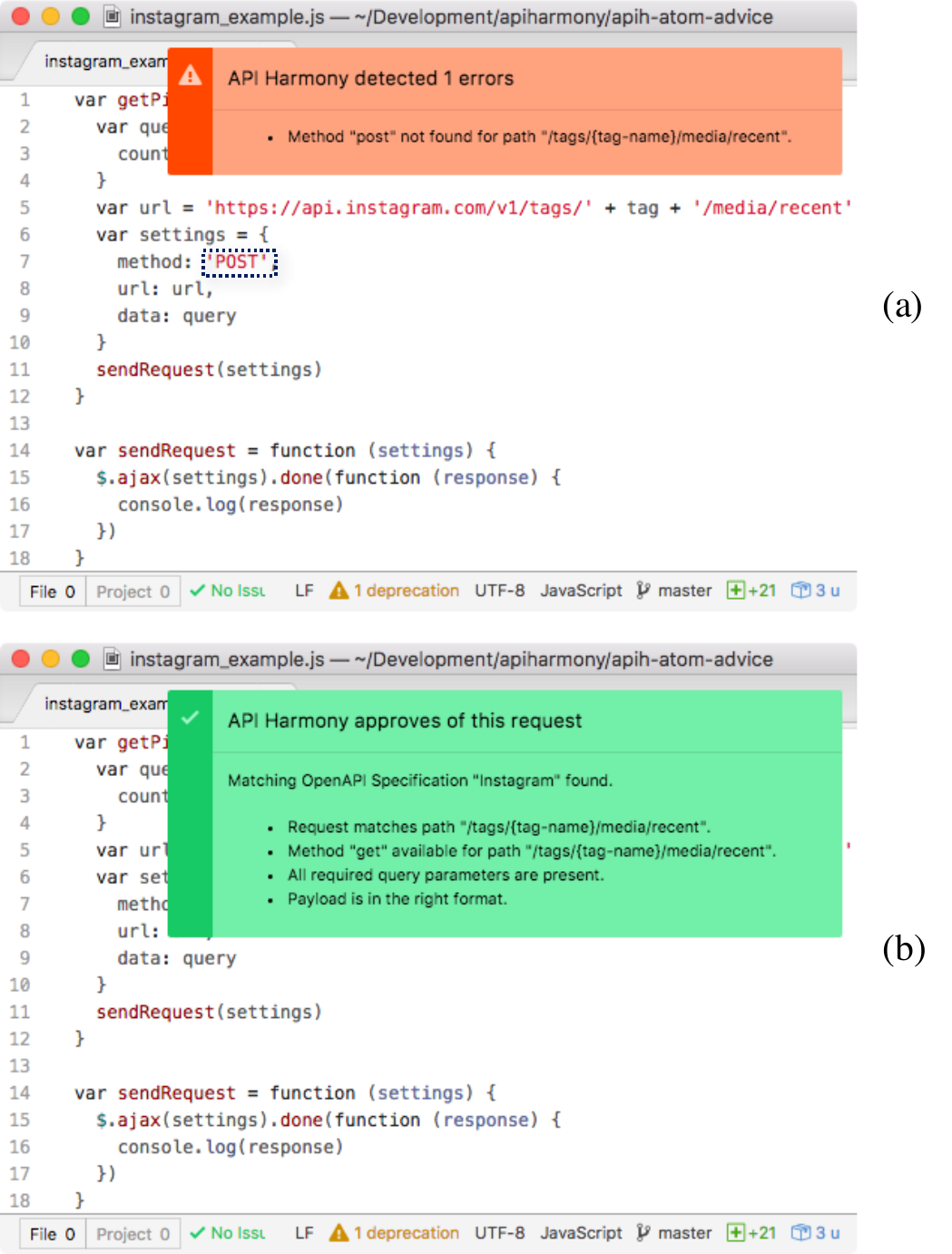}
  \caption{Feedback in Atom (errors highlighted retroactively by spotted boxes): (a) incorrect HTTP method, (b) correct request.}
  \label{fig:atom_screenshots}
  \vspace{-0.2in}
\end{figure}

We implemented a corresponding checking procedure as a prototypical plugin for the Atom code editor~\cite{Atom}. Code checks are performed with some delay upon every user input, given that the edited code contains a web API request. The checking is a relatively expensive operation, which is thus invoked via asynchronous request to be performed by an external service. While this architecture works for an initial prototype, it might need revision due to concerns of sending source code to an external service. Results of the checking procedure are displayed using overlays within the Atom editor, either reporting the error (cf. Figure~\ref{fig:atom_screenshots}a) or what was successfully checked (cf. Figure~\ref{fig:atom_screenshots}b).


\section{Conclusion}
\label{sec:conclusion}
In this paper, we argue for software engineering research addressing the consumption of web APIs -- a topic that is of high relevance in practice, but has seen little attention from research so far.
We discussed the nature of challenges in this field by means of two concrete examples of our work, and presented new work on how research efforts can be applied to support developers in consuming web APIs without errors.


\bibliographystyle{IEEEtran}
\bibliography{IEEEabrv,sigproc}  

\begin{thebibliography}{10}
\providecommand{\url}[1]{#1}
\csname url@samestyle\endcsname
\providecommand{\newblock}{\relax}
\providecommand{\bibinfo}[2]{#2}
\providecommand{\BIBentrySTDinterwordspacing}{\spaceskip=0pt\relax}
\providecommand{\BIBentryALTinterwordstretchfactor}{4}
\providecommand{\BIBentryALTinterwordspacing}{\spaceskip=\fontdimen2\font plus
\BIBentryALTinterwordstretchfactor\fontdimen3\font minus
  \fontdimen4\font\relax}
\providecommand{\BIBforeignlanguage}[2]{{%
\expandafter\ifx\csname l@#1\endcsname\relax
\typeout{** WARNING: IEEEtran.bst: No hyphenation pattern has been}%
\typeout{** loaded for the language `#1'. Using the pattern for}%
\typeout{** the default language instead.}%
\else
\language=\csname l@#1\endcsname
\fi
#2}}
\providecommand{\BIBdecl}{\relax}
\BIBdecl

\bibitem{Fielding:2000}
R.~T. Fielding, ``{Architectural styles and the design of network-based
  software architectures},'' Ph.D. dissertation, U. of California, Irvine,
  2000.

\bibitem{APIHarmony}
{IBM API Harmony}. \url{https://apiharmony-open.mybluemix.net/}.

\bibitem{Mashape}
{PublicAPIs}. \url{https://www.publicapis.com/}.

\bibitem{ProgrammableWeb}
{ProgrammableWeb}. \url{http://www.programmableweb.com/}.

\bibitem{slee2007thrift}
M.~Slee, A.~Agarwal, and M.~Kwiatkowski, ``Thrift: Scalable cross-language
  services implementation,'' \emph{Facebook White Paper}, 2007.

\bibitem{robillard2013automated}
M.~P. Robillard, E.~Bodden, D.~Kawrykow, M.~Mezini, and T.~Ratchford,
  ``Automated {API} property inference techniques,'' \emph{IEEE Transactions on
  Software Engineering}, vol.~39, no.~5, pp. 613--637, 2013.

\bibitem{Suter:2015}
P.~Suter and E.~Wittern, ``{Inferring Web API Descriptions From Usage Data},''
  in \emph{Proc. of the Workshop on Hot Topics in Web Systems and
  Technologies}, 2015.

\bibitem{Wittern:2017}
E.~Wittern, A.~T.~T. Ying, Y.~Zheng, J.~Dolby, and J.~A. Laredo, ``{Statically
  Checking Web API Requests in JavaScript},'' in \emph{Proc. of ICSE, to
  appear}, 2017.

\bibitem{dagenais2011recommending}
B.~Dagenais and M.~P. Robillard, ``Recommending adaptive changes for framework
  evolution,'' \emph{ACM TOSEM}, vol.~20, no.~4, p.~19, 2011.

\bibitem{wu2010aura}
W.~Wu, Y.-G. Gu{\'e}h{\'e}neuc, G.~Antoniol, and M.~Kim, ``Aura: a hybrid
  approach to identify framework evolution,'' in \emph{Proc. of ICSE}, 2010,
  pp. 325--334.

\bibitem{OAI}
{Open API Initiative}. \url{https://openapis.org/specification}.

\bibitem{Bermbach:2016}
D.~Bermbach and E.~Wittern, ``Benchmarking web api quality,'' in \emph{Proc. of
  the International Conference in Web Engineering}, 2016, pp. 188--206.

\bibitem{Li:2014}
D.~Li, S.~Hao, J.~Gui, and W.~G.~J. Halfond, ``An empirical study of the energy
  consumption of android applications,'' in \emph{Proc. of ICSME}, 2014, pp.
  121--130.

\bibitem{Chin:2012}
E.~Chin, A.~P. Felt, V.~Sekar, and D.~Wagner, ``Measuring user confidence in
  smartphone security and privacy,'' in \emph{Proc. of the Symposium on Usable
  Privacy and Security}, 2012.

\bibitem{WADL}
{WADL - Web Application Description Language}.
  \url{http://www.w3.org/Submission/wadl/}.

\bibitem{RAML}
{RAML - RESTful API Modeling Language}. \url{http://raml.org/}.

\bibitem{APIsGuru}
{APIs.guru - Wikipedia for Web APIs}. \url{https://apis.guru/}.

\bibitem{APIConnect}
{IBM API Connect}. \url{https://developer.ibm.com/apiconnect/}.

\bibitem{SwaggerCore}
{Swagger Core Library}. \url{https://github.com/swagger-api/swagger-core}.

\bibitem{SwaggerJSDoc}
{swagger-jsdoc}. \url{https://github.com/Surnet/swagger-jsdoc}.

\bibitem{Sohan:2015b}
S.~M. Sohan, C.~Anslow, and F.~Maurer, ``{SpyREST: Automated RESTful API
  Documentation Using an HTTP Proxy Server},'' in \emph{Proc. of ASE}, 2015,
  pp. 271--276.

\bibitem{Feldthaus_icse2013}
A.~Feldthaus, M.~Sch\"{a}fer, M.~Sridharan, J.~Dolby, and F.~Tip, ``Efficient
  construction of approximate call graphs for {JavaScript IDE} services,'' in
  \emph{Proc. of ICSE}, 2013, pp. 752--761.

\bibitem{DBLP:conf/pldi/SchaferSDT13}
M.~Sch{\"{a}}fer, M.~Sridharan, J.~Dolby, and F.~Tip, ``Dynamic determinacy
  analysis,'' in \emph{Proc. of PLDI}, 2013, pp. 165--174.

\bibitem{DBLP:conf/oopsla/AndreasenM14}
E.~Andreasen and A.~M{\o}ller, ``Determinacy in static analysis for jquery,''
  in \emph{Proc. of OOPSLA}, 2014, pp. 17--31.

\bibitem{DBLP:conf/kbse/KoLDR15}
Y.~Ko, H.~Lee, J.~Dolby, and S.~Ryu, ``Practically tunable static analysis
  framework for large-scale {JavaScript} applications,'' in \emph{Proc. of
  ASE}, 2015, pp. 541--551.

\bibitem{jQueryUsage}
{Usage Statistics of JavaScript Libraries for Websites, August 2016}.
  \url{https://w3techs.com/technologies/overview/javascript_library/all/}.

\bibitem{Atom}
{Atom Editor}. \url{https://atom.io/}.

\end{thebibliography}

\end{document}